\documentclass[12pt]{article}
\usepackage{epsf,psfig,eqsection,cite}

\begin{document}
\setlength{\unitlength}{0.2cm}

\title{Determination of the exponent $\gamma$ for SAWs on the two-dimensional
Manhattan lattice}

\author{
  \\
  {\small Sergio Caracciolo}             \\[-0.2cm]
  {\small\it Scuola Normale Superiore and INFN -- Sezione di Pisa}  \\[-0.2cm]
  {\small\it I-56100 Pisa, ITALY}          \\[-0.2cm]
  {\small Internet: {\tt Sergio.Caracciolo@sns.it}}     \\[-0.2cm]
  \\[-0.1cm]  \and
  {\small Maria Serena Causo}             \\[-0.2cm]
  {\small Peter Grassberger}             \\[-0.2cm]
  {\small\it John von Neumann-Institut f\"ur Computing (NIC)}  \\[-0.2cm]
  {\small\it Forschungszentrum J\"ulich}  \\[-0.2cm]
  {\small\it D-52425 J\"ulich, GERMANY}          \\[-0.2cm]
  {\small Internet: {\tt M.S.Causo@fz-juelich.de}}     \\[-0.2cm]
  {\small Internet: {\tt P.Grassberger@fz-juelich.de}}     \\[-0.2cm]
  \\[-0.1cm]  \and
  {\small Andrea Pelissetto}                          \\[-0.2cm]
  {\small\it Dipartimento di Fisica and INFN -- Sezione di Roma I}  \\[-0.2cm]
  {\small\it Universit\`a degli Studi di Roma ``La Sapienza"}       \\[-0.2cm]
  {\small\it I-00185 Roma, ITALY}          \\[-0.2cm]
  {\small Internet: {\tt pelisset@ibmth.df.unipi.it}}   \\[-0.2cm]
  {\protect\makebox[5in]{\quad}}  % To force authors' names to be written
                                  %   vertically, one above another.
                                  % (\author seems to put them side-by-side
                                  %   if there is room.)
  \\
}
\vspace{0.5cm}

\maketitle
\thispagestyle{empty}   % Suppress page number on front page.

%\ltapprox and \gtapprox produce > and < signs with twiddle underneath
\def\spose#1{\hbox to 0pt{#1\hss}}
\def\ltapprox{\mathrel{\spose{\lower 3pt\hbox{$\mathchar"218$}}
 \raise 2.0pt\hbox{$\mathchar"13C$}}}
\def\gtapprox{\mathrel{\spose{\lower 3pt\hbox{$\mathchar"218$}}
 \raise 2.0pt\hbox{$\mathchar"13E$}}}

\vspace{0.2cm}

\begin{abstract}
We present a high-statistics Monte Carlo determination of 
the exponent $\gamma$ for self-avoiding walks on a Manhattan
lattice in two dimensions. 
A conservative estimate is
$\gamma \gtapprox 1.3425 \pm 0.0003$, in agreement with the
universal value $43/32$ on regular lattices, but in conflict with
predictions from conformal field theory and with a recent estimate 
from exact enumerations. 
We find strong corrections to scaling that seem to indicate the presence 
of a non-analytic exponent $\Delta < 1$. If we assume $\Delta = 11/16$
we find $\gamma = 1.3436 \pm 0.0003$, where the error is purely statistical. 
\end{abstract}
\clearpage

\newcommand{\be}{\begin{equation}}
\newcommand{\ee}{\end{equation}}
\newcommand{\bea}{\begin{eqnarray}}
\newcommand{\eea}{\end{eqnarray}}
\newcommand{\<}{\langle}
\renewcommand{\>}{\rangle}

\newcommand{\R}{\hbox{{\rm I}\kern-.2em\hbox{\rm R}}}

\newcommand{\reff}[1]{(\ref{#1})}

\section{Introduction}

The self-avoiding walk (SAW) is a model 
which describes the universal properties of flexible chain polymers in a 
good solvent in the dilute regime. A simple but intriguing modification
has been recently introduced to study  polymers with an intrinsic orientation
\cite{Miller,Cardy94}.
This orientation could be due to the presence of dipole moments
on the monomers of the chain or to an ordering in the sequence of monomer
constituents.
On the lattice one considers SAWs with a short-range interaction between 
different steps of the walk according to their relative orientation. The 
partition function is simply 
\be
z_N = \sum_{\{\omega\}} e^{\beta_a m_a + \beta_p m_p}\; ,
\ee
where $m_p$ and $m_a$ are respectively the number of parallel and antiparallel
interactions and the sum extends over all SAWs of length $N$.

This model has a rich phase diagram 
\cite{BennetWood95,barkema96,trovato97,prellberg97,Barkema_etal}
and can be analyzed theoretically \cite{Miller,Cardy94} 
by mapping it into a complex $O(n)$ model in the limit $n\to 0$. 
In two dimensions, the theory with a repulsive interaction between
parallel bonds, i.e. with $\beta_a = 0$, $\beta_p < 0$, was analyzed
using conformal invariance techniques \cite{Cardy94}. It was shown 
that the new interaction is truly marginal, giving rise to a line 
of fixed points. 
The main consequence is that the partition-function exponent 
$\gamma$ should vary continuously with the strength of the 
orientation-dependent interaction.  The exponent $\gamma$ is 
defined from the asymptotic behaviour of the partition function
$z_N$, which should scale asymptotically as
\be
z_N = A \mu^N N^{\gamma-1}\,.
\label{cN}
\ee
Here $A$ and $\mu$ are non-universal constants, while
$\gamma$ is an exponent that, in the absence of the orientation-dependent
interaction, is expected to be universal: it
should not depend on the details of the
interaction and it should assume the same value for any two-dimensional
regular lattice. Using Coulomb-gas techniques, Nienhuis \cite{Nienhuis82_84}
predicted $\gamma_{\rm reg} = 43/32$, a value that has been 
confirmed to high-precision by many numerical computations,
see e.g. Ref. \cite{conway93}.
On the other hand, for interacting oriented SAWs with $\beta = 0$, 
$\beta_p < 0$, $\gamma$ should be a function of $\beta_p$.
Unfortunately, conformal field
theory does not provide definite numerical estimates, although
it predicts that $\gamma(\beta_p)$ should decrease monotonically 
as  $\beta_p\to-\infty$.

The square-lattice model was studied by exact enumerations 
in Ref. \cite{BennetWood95} and by transfer-matrix techniques in 
Ref. \cite{koo95} finding a very tiny
dependence of $\gamma$ on $\beta_p$, thereby supporting 
the field-theory analysis.
When parallel interactions are forbidden, i.e. for $\beta_p = -\infty$,
Ref. \cite{BennetWood95} finds 
using unbiased approximants,
\be
0.006 \ltapprox \gamma_{\rm reg}-\gamma \ltapprox  0.013\,.
\ee
The systematic uncertainty due to the extrapolations of two different series is 
taken into account in this range. The evidence for a non-universal behaviour
is not overwhelming, keeping also into account that several problems affected
the results of the analysis. As the authors report, they found small 
shifts of the critical fugacity $\mu_c$ from the value it 
assumes for the ordinary SAW,
in contrast with the theoretical result that 
$\mu_c$ should not depend on $\beta_p$.  More worryingly,
an analysis using biased differential approximants
with fixed critical fugacity  gave a smaller prediction for 
$\gamma_{\rm reg}-\gamma$, although with much less confidence 
since most of the approximants were defective. The analysis of 
Koo \cite{koo95}, based on strips of width $\le 8$, was less precise.
For the largest value of $\beta_p$ that was analyzed, $\beta_p = -3$, he 
finds $\gamma_{\rm reg} - \gamma = 0.018 \pm 0.012$, a difference 
that is barely significant. 

The evidence provided by Refs. \cite{BennetWood95,koo95} for a non-universal
behaviour of $\gamma$ was not conclusive and this spurred many workers 
to improve the result and/or to investigate the problem using different methods.
The transfer-matrix analysis was improved in Ref. \cite{trovato97}: using 
larger strips they did not find any evidence of a non-universal
behaviour and interpreted previous results
as due to short-series effects and to the small size of the strips. 
Other field-theoretical predictions were also tested. 
Refs. \cite{flesia95,barkema96} computed by Monte Carlo and exact-enumeration 
methods the mean value of $m_p$ for ordinary SAWs: they find 
that $\<m_p\>$ converges to a constant as $N\to\infty$, in contrast 
with the field-theoretical prediction $\<m_p\>\sim \log N$. 
Recently the behaviour of $\<m_p\>$ on a cylinder was determined
by a Monte Carlo simulation \cite{Frauenkron-etal}, finding also in
this case a result in disagreement with the
field-theory predictions.

Recently it was shown \cite{Manhattan_Cardy} that
Cardy's original argument implies that the
exponent $\gamma$ for the Manhattan lattice should also be different from the 
exponent $\gamma_{\rm reg}$. 
Indeed on this lattice a SAW is oriented by default and
parallel interactions are automatically suppressed.
{}From the analysis of long exact-enumeration series the authors of
Ref. \cite{Manhattan_Cardy} report
\be
\gamma_{\rm reg} - \gamma = 0.0053 \pm 0.0030 \,.
\label{stima}
\ee
The effect is extremely small. It differs from zero by less than two
error bars. The evidence for 
$\gamma \ne \gamma_{reg}$ is therefore not overwhelmingly persuasive,
and the theoretical importance of the problem asks for further investigations. 
Indeed one can suspect that the small deviation is simply a systematic effect
due to the corrections to scaling that are not completely taken into account
by the analysis.

We have therefore decided to investigate the problem by means of a Monte Carlo 
simulation, computing the exponent $\gamma$ for SAWs on a Manhattan lattice.
The advantage is that we are able to work with very long walks 
($N\le N_{\rm max} = 32000$)
and therefore to reduce the unknown 
systematic uncertainty due to the extrapolation
$N \to \infty$.

Our simulations were performed with two different algorithms.
The first one, the {\em join-and-cut} algorithm  \cite{join-and-cut},
is a dynamic Monte Carlo algorithm
that  works in the ensemble of couples of walks with fixed total length.
The algorithm is at present the best one to compute the 
exponent $\gamma$ since
the autocorrelation time in CPU units scales as
$N^{1.6}$, while for other algorithms it behaves no better than 
$N^{\approx 2}$.
The second algorithm is a variant of the pruned-enriched Rosenbluth 
method (PERM) \cite{Grassberger_97}.
This is a growth algorithm. Asymptotically, it is slower than the 
{\em join-and-cut} algorithm in generating independent configurations. 
The computer time to generate an independent configuration scales as 
$N^2$. But the constant in front of $N^2$ is very small, for instance the 
number of monomer additions needed to obtain one independent configuration
was numerically found to increase as
 $\approx 0.008 N^2$ for $N>10000$. Therefore
it is possible that the PERM is the most efficient one even for
quite long walks. As we shall discuss, with a clever improvement,
the {\em Markovian anticipation}, the PERM is more efficient
in providing estimates of $\gamma$ than the join-and-cut algorithm
as long as the length of the sampled walks is less than $10^4$.
Only for longer walks the join-and-cut algorithm is faster.
Another advantage in the
present context is that it gives directly, together with the estimate 
of the partition sum for chains of length $N_{\rm max}$ and without 
extra cost, also all partition sums for shorter chains.
These estimates for different $N$ are not independent, but just 
because of this fact they are particularly useful for estimating $\gamma$.

The main sources of systematic errors in our analysis are the corrections to 
scaling, assumed to be of the form
\be
{z_N \over A \mu^N N^{\gamma -1}} \approx 1+{a \over{N^\Delta}}+{a_1 \over N^{\Delta_1}}
+ \cdots
\ee 
with $\Delta < \Delta_1 < \cdots$.
All numerical evidence for SAWs on
regular lattices (square, honeycomb and triangular)
indicates that the leading correction is the 
analytic one \cite{Guttmann_88,Conway_96,subleading}.
On the other hand Saleur \cite{Saleur_87} predicted $\Delta = 11/16$,
a result that was confirmed in numerical work on lattice 
trails \cite{Conway_93,Guim_96}.\footnote{An 
additional hint to $\Delta = 11/16$ is 
the numerical observation of Barkema and Flesia \cite{barkema96} that 
the average number of loops of length $l$ forming a parallel contact 
scales as $\langle m_p\rangle_l \sim l^{-1.65\pm 0.05}$. From this they estimate
that the total number of parallel contacts behaves as 
$\langle m_p\rangle = a - b/N^{0.65}$. On the other hand, $\langle m_p\rangle$ 
is proportional to $dz_N(T)/dT$ if one includes an orientation 
dependent interaction, whence there should be generically a term 
$\sim N^{-0.65}\approx N^{-11/16}$ in $z_N(T)$.}
Why this exponent does not show up for SAWs on regular lattices is 
completely unclear. 

For the Manhattan lattice, Ref. \cite{Manhattan_Cardy} showed that 
the enumeration data are very well fitted using an Ansatz 
with no non-analytic terms\footnote{Inclusion 
of a term with $\Delta = 11/16$ worsens the quality of the 
extrapolation \cite{Tony_private}.}
with $\Delta < 1$. On the other 
hand a naive fit of our Monte Carlo data would indicate just the opposite:
the results are well fitted assuming a correction-to-scaling exponent
of order $\Delta \approx 0.5-0.7$. Of course one should not take this
indication too seriously --- our data are not precise enough for a 
serious attempt to determine $\Delta$ --- but it is fair to say 
that $\Delta = 11/16$ is our preferred value. If we assume 
$\Delta = 11/16$ we obtain
\be
\gamma = 1.3436 \pm 0.0003,
\label{gamma11su16}
\ee
in very good agreement with $\gamma = \gamma_{\rm reg} = 43/32 = 1.34375$.
We have also tried to analyze our data assuming $\Delta = 1$. 
The quality of the fit is somewhat worse, although one could think 
that this is simply due to the additional neglected corrections to 
scaling\footnote{A similar phenomenon occurs for SAWs on the square lattice
\cite{subleading}. 
If one analyzes the end-to-end distance for short SAWs, with a single correction term,
one finds $\Delta \approx 0.80$. Only the inclusion of longer walks  and
more correction terms with $\Delta_i > 1$ 
gives $\Delta \approx 1.0$.} that are still relevant 
at the values of $N$ we are working (the determination of $\Delta$
depends mainly on the data with $N_{\rm tot} = 2000$, i.e. on walks with 
$N\ltapprox 1000$).

We do not know how reliable the estimate
\reff{gamma11su16} is and a serious analysis of the systematic errors
is practically impossible. In any case, without explicit assumptions
on $\Delta$, we can still obtain the lower bound
\be
\gamma \gtapprox 1.3425 \pm 0.0003.
\ee
Although this result is lower than the estimate \reff{gamma11su16}, 
it clearly supports the fact that $\gamma = \gamma_{\rm reg}$.
The prediction of Ref. \cite{Manhattan_Cardy}, 
$\gamma = 1.3385 \pm 0.003$ is instead clearly excluded.

We should point out that if we analyze only data with small values of 
$N$, we would obtain 
a lower estimate for $\gamma$, in close agreement with 
the result \reff{stima}.
This shows the crucial role played in this problem by corrections to 
scaling and the importance of performing simulations 
for very large values of $N$. 

\section{The join-and-cut algorithm on the Manhattan lattice}

\subsection{Description of the algorithm}

In this Section we will define the pivot and the join-and-cut algorithm on a 
Manhattan lattice. The Manhattan lattice is a two-dimensional 
square lattice on which bonds are directed in such a way that adjacent rows 
(columns) have antiparallel directions, corresponding to the 
traffic pattern in Manhattan. Although bonds are directed, there 
is no overall directional bias. Explicitly we will assume
the following orientations: a vertical bond connecting the 
points of coordinates $(x,y)$, $(x,y+1)$ is directed upward if
$x$ is even, downward if $x$ is odd; a horizontal bond 
connecting the points of coordinates $(x,y)$ and $(x+1,y)$ is 
directed to the left if $y$ is even, to the right if $y$ is odd.

Let us now define the pivot algorithm 
\cite{Lal,MacDonald,Sokal1}. It works in the ensemble 
of fixed-length walks with free endpoints: we will be interested in
self-avoiding walks, 
but the algorithm can also be applied 
in a very efficient way to the Domb-Joyce model
\cite{Parisi,Nickel}, to power-law walks
\cite{Parisi}, and to interacting polymers far from the 
$\Theta$-transition \cite{zifferer,canadesi}.

The algorithm works as follows \cite{Sokal1}. 
Given an $N$-step SAW $\omega$ starting at 
the origin and ending anywhere, 
$\omega\equiv\{\omega(0),\ldots,\omega(N)\}$, an iteration of the 
algorithm consists of the following steps:
\begin{enumerate}
\item[(i)] choose randomly, with uniform probability, an integer 
$k\in\{0,1,\ldots,N-1\}$;
\item[(ii)] choose with probability $P(g)$ an element $g$ of the 
lattice point symmetry group. The probability must 
satisfy $P(g) = P(g^{-1})$ to ensure detailed balance;
\item[(iii)] propose a pivot move $\omega\to\omega'$ defined by
\be
{\omega'}(i) = \cases{\omega(i) & for $1\le i \le k$, \cr
    \omega(k) + g(\omega(i) - \omega(k)) & for $k+1\le i \le N$.}
\ee
The proposed move is accepted if $\omega'$ is self-avoiding;
otherwise it is rejected and we stay with $\omega$.
\end{enumerate}
The probability $P(g)$ must be such to ensure ergodicity:
as discussed in Refs. \cite{Sokal1,Madras90} not all symmetry
transformations are needed. In particular \cite{Madras90} the pivot algorithm
is ergodic on a square lattice if $P(g)$ is non-vanishing only for 
diagonal reflections, that is for reflections through lines of 
slope $\pm 1$. 

For the Manhattan lattice the point symmetry group is much smaller than
the symmetry group of the square lattice. Indeed the lattice is symmetric
only with respect to lines of slope $-1$ going through lattice points 
$(x,y)$ with $x+y$ even, and with respect to lines of slope 
$+1$ going through $(x,y)$ with $x+y$ odd. Therefore we will
modify step (ii) in the following way:
\begin{enumerate}
\item[(ii)] 
if $\omega_x(k) + \omega_y(k)$ 
is even (resp. odd), let $g$ be the reflection with 
respect to a line of slope $-1$ (resp. $+1$), going through 
$\omega(k)$.
\end{enumerate}
Since $g^2=1$, detailed balance is automatically satisfied. The 
tricky point is ergodicity. It can be proved by slightly 
modifying the proof of Section 5 of Ref. \cite{Madras90}. 
Using the same definitions, the basic observation is the following:
if $\omega(k)\equiv (x_k,y_k)$ is a walk site belonging to
a diagonal support line of slope $-1$ (resp. $+1$), then
$x_k+y_k$ is even (resp. odd). Using this fact the proof works without 
any change.

We want now to define the join-and-cut algorithm \cite{join-and-cut}. 
The tricky point here
is that in general, given two walks on the Manhattan lattice,
their concatenation is a walk that does not respect the bond orientation 
of the lattice. More precisely, given two walks $\omega_1$, $\omega_2$
of lengths $N_1$ and $N_2$, the 
concatenated walk $\omega=\omega_1\circ\omega_2$ respects the orientation
of the lattice only if\footnote{The only exception to this rule is  
when $\omega_2$ is a rod.}
\be
{\rm mod}(\omega_{1x}(N_1) - \omega_{2x}(0),2) = 0, \qquad
{\rm mod}(\omega_{1y}(N_1) - \omega_{2y}(0),2) = 0.
\ee
This reflects the fact that only translations of two steps in each direction
are symmetries of the lattice. A second consequence of the bond orientation 
of the Manhattan lattice is the following: consider a walk $\omega$
and cut it into two parts $\omega_1$ and $\omega_2$ such that 
$\omega = \omega_1 \circ \omega_2$. In general there is no
lattice translation $T$ such the translated walk $T\omega_2$ 
respects the bond orientation of the lattice and satisfies
$(T\omega_2)(0) = \omega_1(0) = \omega(0)$. 
Indeed this happens only if 
\be
{\rm mod}(\omega_{1x}(0) - \omega_{2x}(0),2) = 0, \qquad
{\rm mod}(\omega_{1y}(0) - \omega_{2y}(0),2) = 0.
\ee
With these two observations in mind we define our ensemble in the following
way: $T_{N_{\rm tot}}$ consists of all pairs $(\omega_1,\omega_2)$ of SAWs, 
each walk starting either in $(0,0)$ or in $(1,1)$ and ending anywhere,
such that the {\em total} number of steps in the two
walks is some fixed {\em even} number $N_{\rm tot}$. Moreover we require 
the lengths of $\omega_1$ and $\omega_2$ to be even. 
Explicitly
\be
T_{N_{\rm tot}} = \bigcup_{k=1}^{N_{\rm tot}/2-1}
   (S_{2k}(0,0) \cup S_{2k}(1,1)) \times 
   (S_{N_{\rm tot} - 2k}(0,0) \cup S_{N_{\rm tot} - 2k}(1,1))
\label{ensemble}
\ee
where $S_k(x,y)$ is the set of $k$-step walks starting from $(x,y)$ and 
ending anywhere.
Each pair in the ensemble is given
equal weight: therefore the two walks are not interacting except for the 
constraint on the sum of their lengths. Notice that there is a 
one-to-one correspondence between $S_k(0,0)$ and $S_k(1,1)$ so that the 
ensemble defined in Eq. \reff{ensemble} is equivalent to 
\be
T_{N_{\rm tot}} = \bigcup_{k=1}^{N_{\rm tot}/2-1}
 S_{2k}(0,0)  \times S_{N_{\rm tot} - 2k}(0,0) 
\ee
One sweep of the algorithm consists of the following steps:
\begin{itemize}
\item[(i)] Starting from a pair of walks $(\omega_1,\omega_2)$, 
we update each of them independently using some 
ergodic fixed-length algorithm. We use the pivot 
algorithm we described above.
\item[(ii)] With probability $1/2$ we interchange 
$\omega_1$ and $\omega_2$, i.e. 
$(\omega_1,\omega_2)\to(\omega_2,\omega_1)$.
\item[(iii)] We first check the parity of the endpoints:  if 
${\rm mod}(\omega_{1x}(N_1) - \omega_{2x}(0),2) = 1$, 
we stay with $(\omega_1,\omega_2)$ . Otherwise we attempt 
a {\em join-and-cut} move.  We choose with 
uniform probability $k\in\{1,\ldots,N_{\rm tot}/2-1\}$. 
Then we concatenate the two walks
$\omega_1$ and $\omega_2$ forming a new (not necessarily 
self-avoiding) walk $\omega_{\rm conc}=\omega_1\circ\omega_2$; 
then we cut $\omega_{\rm conc}$ 
creating two new walks ${\omega'}_1$ and ${\omega'}_2$ of lengths 
$2k$ and $N_{\rm tot}-2k$. 
If ${\omega'}_1$ and ${\omega'}_2$ are self-avoiding we keep them;
otherwise the move is rejected and we stay with $\omega_1$ and $\omega_2$.
\end{itemize}
It is easy to see that the full algorithm is ergodic.

The algorithm defined on the Manhattan lattice works essentially 
in the same way as the standard one. The only important difference 
is that one must check the parities of $\omega_1(N_1)$ and $\omega_2(0)$
before attempting the join-and-cut move. This check is successfull 
in 50\% of the cases and thus the algorithm we have defined above 
should be essentially equivalent to the standard algorithm 
in which one performs two pivot updates of each walk for every 
join-and-cut move (in the notation of Ref. \cite{join-and-cut}
it corresponds to the algorithm with $n_{\rm piv} = 2$). 
In principle it is easy to avoid this 50\% rejection by modifying the 
third step of the algorithm in the following way:
\begin{itemize}
\item[(iii)] {\em [improved join-and-cut move]}.
If 
${\rm mod}(\omega_{1x}(N_1) - \omega_{2x}(0),2) = 1$, 
let $\omega_{\rm conc} = \omega_1 \circ R\omega_2$, where $R\omega_2$
is the walk reflected with respect to the line of slope +1 
going through $\omega_2(0)$; if
${\rm mod}(\omega_{1x}(N_1) - \omega_{2x}(0),2) = 0$, 
let $\omega_{\rm conc} = \omega_1 \circ \omega_2$. Then proceed as before.
\end{itemize}
This modification attempts twice the number of join-and-cut moves
with respect to the previous one, and thus it should be more efficient.
However the difference in performance is not expected to be large. 
Indeed from the analysis of the autocorrelation times of Ref. 
\cite{join-and-cut}, we obtain\footnote{\label{tauint-npiv2} 
Notice that 
in Ref. \protect\cite{join-and-cut} $\tau_{\rm int}(n_{\rm piv})$
is reported in units of two iterations, each iteration consisting 
of $n_{\rm piv}$ pivot attempts and one join-and-cut attempt. In order to 
make a correct comparison we should multiply $\tau_{\rm int}(n_{\rm piv})$ 
by $n_{\rm piv}$. Thus what we report as $\tau_{\rm int}(n_{\rm piv}=2)$ corresponds 
to $2 \tau_{\rm int}(n_{\rm piv}=2)$ of Ref. \cite{join-and-cut}.} that the ratio
$\tau_{\rm int}(n_{\rm piv} = 2)/ \tau_{\rm int}(n_{\rm piv} = 1)$ is equal to 
$1.6,1.5,1.2$ for $N_{\rm tot} = 500,2000,8000$ respectively. 
For the algorithm on the Manhattan lattice we expect 
$\tau_{\rm int}({\rm non\!-\!impr})/\tau_{\rm int}({\rm imp})$ to be very similar
to the previous ratio: therefore for $N_{\rm tot} = 8000$ we expect that 
the autocorrelation time for the original algorithm to be only
20-30\% higher than the corresponding time for the improved one. 
One should also keep in mind that one iteration of the 
improved algorithm is slower than one iteration of the 
unimproved one, since one is doing twice the number of join-and-cut attempts.
Therefore we expect that for $N_{\rm tot}=8000$ the improved algorithm 
is only 10-20\% better than the original one.
Since the implementation of the 
improved algorithm required major changes of our codes, we decided
to work with the unimproved version we described above.

\subsection{Determination of $\gamma$ from join-and-cut data} \label{sec2.2}

Let us now discuss how the critical exponent $\gamma$ can be estimated from 
the Monte Carlo data produced by the join-and-cut algorithm.

Let us start by noticing that the random variable $N_1$, the length of the 
first walk, has the distribution
\be
{\overline\pi}(N_1) \, =\, \cases{
 \displaystyle{z_{N_1} z_{N_{\rm tot}-N_1}\over Z(N_{\rm tot})} & if $N_1$ is even, \cr
 \displaystyle{\vphantom{1\over2}0}                    & if $N_1$ is odd} 
\label{pi-bar}
\ee
for $1\le N_1\le N_{\rm tot}-1$; here $Z(N_{\rm tot})$ is the obvious 
normalization factor and $z_N$ is the number of $N$-step walks going from the 
origin to any lattice point whose asymptotic behaviour 
for large $N$ is given by \reff{cN}.
The idea is then to make inferences of $\gamma$ from the observed 
statistics of $N_1$. Of course the problem is that \reff{cN} is an 
asymptotic formula valid only in the large-$N$ regime. We will thus 
proceed in the following way: we will suppose that \reff{cN} is valid
for all $N\ge N_{\rm min}$ for many increasing values of $N_{\rm min}$ 
and correspondingly we will get estimates 
$\hat{\gamma}(N_{\rm tot},N_{\rm min})$; 
these quantities are {\em effective} exponents that depend on $N_{\rm min}$ and 
that give correct estimates of $\gamma$ as $N_{\rm min}$ and 
$N_{\rm tot}$ go to infinity.

The determination of $\gamma$ from the data is obtained using the 
maximum-likelihood method (see e.g. \cite{Silvey}). 
We will present here only the 
results: for a detailed discussion we refer the reader to 
\cite{join-and-cut}.

Given $N_{\rm min}$, consider the function (from now on 
we suppress the dependence on $N_{\rm tot}$)
\be
\theta_{N_{\rm min}} (N_1) =\, 
   \cases{1 & if $N_{\rm min}\le N_1\le N_{\rm tot} - N_{\rm min}$, \cr
          0 & otherwise,}
\ee
and let $X$ be the random variable 
\be
X = \log[ N_1(N_{\rm tot}-N_1)] \;\; .
\ee
Then define
\be
X^{\rm cens}(N_{\rm min}) = {\< X\theta_{N_{\rm min}}\> \over \< \theta_{N_{\rm min}}\>}
\label{Xcens}
\ee
where the average $\<\ \cdot\ \>$
is taken in the ensemble $T_{N_{\rm tot}}$ sampled 
by the join-and-cut algorithm. The quantity defined in Eq. \reff{Xcens}
is estimated in the usual way from the Monte Carlo data obtaining in this 
way $X^{\rm cens}_{\rm MC} (N_{\rm min})$.
Then $\hat{\gamma}(N_{\rm min})$ is computed by solving the equation
\be
X^{\rm cens}_{\rm MC} (N_{\rm min}) \, =\, [X]_{{\rm th},\hat{\gamma}} (N_{\rm min}),
\label{eq_max_like}
\ee
where, for every function of $N_1$, we define
\be
[f(N_1)]_{{\rm th},\gamma} (N_{\rm min}) \equiv 
  {\sum_{N_1 = N_{\rm min}}^{N_{\rm tot}-N_{\rm min}} 
      f(N_1) N_1^{\gamma-1} (N_{\rm tot}-N_1)^{\gamma-1} \over 
   \sum_{N_1 = N_{\rm min}}^{N_{\rm tot}-N_{\rm min}} 
       N_1^{\gamma-1} (N_{\rm tot}-N_1)^{\gamma-1} };
\label{mean-value-theor}
\ee
here the sum is extended over {\em even} values of $N_1$.
The variance of $\hat{\gamma}(N_{\rm min})$ is then given by
\be
{\rm Var}\ [\hat{\gamma}(N_{\rm min})] = \
   {{\rm Var}\ (X^{\rm cens}_{\rm MC} (N_{\rm min}))\over 
    ([X;X]_{{\rm th},\hat{\gamma}} (N_{\rm min}))^2 },
\ee
where $[X;X] = [X^2] - [X]^2$.
We must finally compute ${\rm Var}\ (X^{\rm cens}_{\rm MC} (N_{\rm min}))$.
As this quantity is defined as the 
ratio of two mean values (see Eq. \reff{Xcens}) one must take into account
the correlation between denominator and numerator. 
Here we have used the standard formula for the variance of a ratio
(valid in the large-sample limit)
\be
{\rm Var}\left({A\over B}\right)\, =\, 
   {\<A\>^2\over \<B\>^2} {\rm Var} 
  \left( {A\over \<A\>} - {B\over \<B\>} \right).
\label{varformula}
\ee

\subsection{Dynamic behaviour}

Let us first discuss the simulations using the join-and-cut 
algorithm. We have performed high-statistics runs
at $N_{\rm tot}=500,2000,8000$ and $32000$. 
In  Table \ref{table_iter} 
we report the number of iterations and the CPU time per iteration
on an Alpha-Station 600 Mod 5/266.
The total join-and-cut runs took $1$ year of CPU on
this machine.

In Table \ref{table_tau} we report the acceptance fraction 
of the pivot and of the join-and-cut moves, and, 
for two different values of $N_{\rm min}$,
the autocorrelation times for the observable
\be
Y(N_{\rm min}) = {X\theta_{N_{\rm min}}\over \<X\theta_{N_{\rm min}}\>} -
           {\theta_{N_{\rm min}}\over \<\theta_{N_{\rm min}}\>}
\ee
that, according to Eq. \reff{varformula}, controls the errors on $\gamma$.
Notice that $Y(2) = X/\<X\>-1$, so that 
$\tau_{{\rm int},Y(2)} = \tau_{{\rm int},X}$. 

To compute the autocorrelation times we used the recipe of 
Ref. \cite[Appendix C]{Sokal2}. Indeed the autocorrelation 
function has a very long tail with a very small amplitude, due to the 
fact that $\tau_{\rm exp} \gg \tau_{{\rm int},Y}$. For this reason the 
self-consistent windowing method of 
Ref. \cite[Appendix C]{Sokal1} does not work correctly 
and underestimates the autocorrelation time. 
Following Ref. \cite{Sokal2}, we have computed $\tau_{{\rm int},Y}$ as
\be
\tau_{{\rm int},Y} = \overline{\tau}_{{\rm int},Y} + \tau_{{\rm tail},Y}.
\ee
Here $\overline{\tau}_{{\rm int},Y}$ is the autocorrelation time 
computed using the self-consistent windowing method of 
Ref. \cite[Appendix C]{Sokal1} with window $W = 15 \overline{\tau}_{{\rm int},Y}$
and 
\be
\tau_{tail,Y} = \rho_Y(W) W \log\left( {N\over W f_{\rm piv}}\right)
\label{tautail}
\ee
where $\rho_Y(t)$ is the normalized autocorrelation function and 
$f_{\rm piv}$ is the acceptance fraction of the pivot algorithm.
Of course Eq. \reff{tautail} is a very rough {\em ad hoc} 
estimate of the contribution 
of the tail.
For $X$, 
it amounts to approximately 35\% for $N_{\rm tot} = 32000$, 24\% for $N_{\rm tot} = 8000$,
15\% for $N_{\rm tot}=2000$, and 5\% for 
$N_{\rm tot}=500$. 
Based on our experience with the pivot algorithm we have assigned to 
$\tau_{tail,Y}$ an error of 10\%. The error on $\overline{\tau}_{{\rm int},Y}$
was instead computed as in Ref. \cite{Sokal1}.

We have first analyzed the acceptance fractions. The general analysis of 
Ref. \cite{join-and-cut} predicts $f_{\rm piv} \sim N^{-p}$ and 
$f_{\rm jc} \sim N^{-q}$, 
with $p\approx 0.19$ and $q\approx \gamma - 1\approx 0.34$.
A least-square fit to the data of Table \ref{table_tau} gives
\bea
p &=& 0.19605 \pm 0.00001 \qquad \chi^2_1 = 603, \\
q &=& 0.32522 \pm 0.00004 \qquad \chi^2_1 = 9197.
\eea
The very large value of $\chi^2$ indicates that the quoted errors on
$p$ and $q$, that are of purely statistical nature, are unreliable.
Indeed a simple power law does not fit the data at this level of
precision.
A more realistic estimate of the errors would be
\be
p = 0.1960 \pm 0.0002, \qquad q =  0.325 \pm 0.004.
\ee
We have then performed an analogous analysis for the autocorrelation time 
$\tau_{{\rm int},Y(N_{\rm min})}$. A least-square fit of the form
\be
\tau_{{\rm int},X} \sim N^z 
\ee
gives
\be
z =\, \cases{0.847 \pm 0.006 \qquad \chi^2_1 = 71 & for $N_{\rm min} = 1$, \cr
             0.672 \pm 0.004 \qquad \chi^2_1 = 280 & for $N_{\rm min} = 100$, \cr
             0.747 \pm 0.003 \qquad \chi^2_1 = 50 & for $N_{\rm min} = 200$.}
\ee
Again, the very large value of $\chi^2$ indicates that the error bars are
underestimated by at least a factor of ten.
The results are  somewhat higher than the estimate reported in 
Ref. \cite{join-and-cut}, $z = 0.70 \pm 0.03$. 
This is evident if one directly compares 
our results of Table \ref{table_tau} with the values of 
$\tau_{\rm int}$ reported in Ref. \cite{join-and-cut} 
(see footnote \ref{tauint-npiv2}): 
$\tau_{\rm int} = 4.38(6)$, $9.47(21)$, $20.0(6)$ for $N_{\rm tot} = 500,2000,8000$ 
respectively.
For $N_{\rm tot} = 500$ the autocorrelation times are similar, while 
for $N_{\rm tot} = 8000$ there is a factor-of-two difference. It is very
difficult to believe that the join-and-cut algorithm has a different
dynamic behaviour on the square lattice and on the Manhattan lattice.
We have thus tried to understand if the authors of Ref. \cite{join-and-cut}
underestimated the autocorrelation times. Therefore we have determined 
$\tau_{\rm int}$ from our data using the self-consistent windowing procedure 
with a window of $5\tau_{{\rm int},X}$ as done in Ref. \cite{join-and-cut}.
We obtain 
$\tau_{\rm int} = 4.166(2)$, $10.093(6)$, $29.13(2)$, $98.0(2)$ for $N_{\rm tot} = 500$,
$2000$, $8000$, $32000$
respectively. A fit of these results gives $z = 0.7271(2)$.
This is now in good agreement with the results of Ref. \cite{join-and-cut}.

In conclusion we believe a fair estimate would be 
\be
z=0.75 \pm 0.05\,.
\ee
%\medskip

As noticed in Ref. \cite{join-and-cut}, with a clever implementation of 
the algorithm, it is possible that the CPU-time per iteration 
increases as $N^\sigma$ with $\sigma<1$. In particular it was 
predicted that $\sigma= 1-p$ where $p$ is the exponent 
that controls the pivot acceptance fraction. In two dimensions
we expect $p = 0.1953(21)$ \cite{Sokal1} and this is confirmed 
by our data on the acceptance fraction. From the data of 
Table \ref{table_tau} we obtain 
\be
\sigma = 0.865(1) \qquad \chi^2_1 = 55
\ee
Keeping into account that the error 
bars are underestimated, we find reasonable
agreement with the prediction $\sigma= 1-p$.
Of course what one is really interested in is the autocorrelation
times expressed in CPU units.

We find
\be
\tau_{\rm int} \sim N^{z+\sigma} \sim N^{1.6}\,.
\ee
The algorithm is not optimal, but still represents an improvement with
respect to other
algorithms that behave as $N^{\approx 2}$.

\section{PERM} 

\subsection{Description of the algorithm}

The pruned-enriched Rosenbluth method (PERM) is a chain-growth algorithm based 
on the well-known Rosenbluth (inverse restricted sampling) \cite{Madras-Slade} 
algorithm. In the latter, chains are grown step by step. In unbiased sampling, 
steps are randomly generated, and, if a new step violates the self-avoidance 
constraint, the configuration is discarded and one restarts from scratch.
This leads to exponential `attrition', 
i.e. the number of attempts needed to generate a chain of length $n$ increases 
exponentially with $n$. 

In Rosenbluth sampling, such steps are replaced (if possible) by 
`correct' steps. This diminishes the attrition problem, 
but it leads to a bias that has to be compensated by a weight factor. 
If $m_n$ is the number of free sites where to place the $n$-th monomer (i.e., 
the number of allowed steps at time $n$), then the weight of a chain of length 
$N$ is $W_N \propto \prod_{n=1}^N m_n$. The proportionality factor depends 
on the ensemble one wants to simulate, e.g. it is constant for the canonical 
ensemble with fixed fugacity. The main problem with the Rosenbluth method is the 
fact that these weights show large fluctuations for large $N$. Thus even large 
samples can be completely dominated by just a few events, which leads to large 
statistical errors. Even worse, for very 
large $N$, the part of the distribution of weights which should dominate might 
not have been sampled at all, and this can lead to systematic errors. More 
precisely, while the method is guaranteed to give an unbiased estimate of the 
partition sum, due to these fluctuations and to the convexity of the logarithm, 
estimates of free energies are systematically too small.

The main idea in PERM \cite{Grassberger_97} is to watch the weight $W_N$ as $N$ is growing, 
and to compare it with the estimated average weight for this value of $N$. If 
$W_N$ is judged too big, a clone of the present configuration is made, both
clones are attributed a weight $W_N/2$, and both are independently continued 
(`enrichment'). On the other hand, if $W_N$  is too small, a (pseudo-) random 
number $r$ is drawn uniformly from $[0,1]$. If $r<1/2$, the configuration is 
abandoned (`pruning'), otherwise it is kept and its weight 
is doubled. This 
is most easily implemented by recursive function calls. A pseudocode is given
in \cite{Grassberger_97}.

In this way the problem of large weight fluctuations is avoided, 
although not entirely. 
The set of generated configurations now is correlated due to cloning. Let us 
call a {\it tour} a set of configurations obtained 
by cloning from the same `ancestor'. Although the weights 
of individual configurations vary between narrow limits, the 
weights of tours can fluctuate very much. If so, we are back at basically the 
same problem. This can be avoided by making additional 
biases. For instance, before 
each step one might look ahead $k$ steps, and choose the direction accordingly
\cite{meirovitch}. This is rather time consuming for large $k$, nevertheless it 
can be efficient for dense (collapsed) systems. 

For diluted systems, as it is the case here, the efficiency can be improved  
using 
{\it Markovian anticipation} \cite{Frauenkron-etal,markov}. Here 
one keeps the last $k$ steps in memory, and biases the next step on the 
basis of the statistics of ($k+1$)-step 
configurations accumulated during a previous run. 

On the Manhattan lattice, an $N$-step walk with fixed starting point and fixed 
direction of the first step can be encoded as 
a binary sequence of length $N-1$. A straight step is encoded as `0', 
while a bend by $\pm 90^{\rm o}$ is encoded as `1'. 
Notice that one does not have 
to specify the direction of the bend, as only one bend is allowed at any time 
step. If the starting point has even $x$ and $y$ and the first step is upward, 
then the first bend is to the right (left) if it occurs at even (odd) times. 
After that, each bend is in the same (opposite) 
direction as the previous one if the
number of in-between straight steps is even (odd). 
Let ${\bf S}=(s_{-k},\ldots,s_0) 
\equiv ({\bf s},s_0)$ denote a binary string of length $k+1$, and let 
${\cal C}_{N,m}({\bf S})$ be the 
{\it cylinder set} of all $N$-step walks starting 
upward from $(x,y)=(0,0)$, for which steps 
$m-k,\ldots,m$ are given by ${\bf S}$.
The total weight of this set in an unbiased ensemble is denoted by 
$W_{N,m}({\bf S})$. Finally, we consider 
\be
   p_{N,m}(s_0|{\bf s}) = {W_{N,m}({\bf S})\over \sum_{s_0' = 0,1} W_{N,m}({\bf s},s_0')} \;.
                 \label{markov-a}
\ee
If SAW's were a $k$-th order Markov process, we would obtain perfect importance 
sampling for $N$-step walks if we would select the $m$-th step according to 
$p_{N,m}(s_0|{\bf s})$. Given the fact that SAW's are not Markovian, 
Eq.~(\ref{markov-a}) comes closest to 
importance sampling given a finite ($k$-step) memory. 

In practice, Eq.~(\ref{markov-a}) is not applicable as it stands, since it is 
practically impossible to acquire and store the needed statistics for all $m$ and $N$. 
First of all we use the fact that $p_{N,m}(s_0|{\bf s})$ becomes independent of 
$N$ and $m$ in the limit $m,N-m \gg k$. 
We therefore estimate it for large values of $m$ and $N$ by accumulating 
statistics for all $N$ larger than some lower threshold (we used $N>400$), and 
$N-m$ fixed (we used $N-m = 200$). The fact that this $p(s_0|{\bf s})$ does not 
give correct importance sampling for $m\approx N$ is not serious and does 
not reduce much the efficiency. More serious is the fact 
that it would be inappropriate for small $N$ and $m$.
In particular, for $m<k$ we have no 
string ${\bf s}$ to condition upon. We dealt 
with this problem by tapering the symbol sequence with a string of $k$ zeroes, 
and by taking step $s_0$ in the $m$-th step with probability
\be
   p_m(s_0|{\bf s}) = 
       {const/m + \lim_{m,N-m\to\infty} p_{N,m}(s_0|{\bf s}) \over 
       \sum_{s_0' = 0,1} [const/m + \lim_{m,N-m\to\infty} p_{N,m}(s_0|{\bf s})] }\;. 
\ee
with $const \approx 20$. 

Notice that all this fiddling is not crucial. The method 
{\em per se} is correct 
even if the probabilities for taking the $m$-th step were chosen badly, 
provided the probability is not set equal to zero for any allowed step.
But efficiency can increase substantially with a good choice. Efficiency can 
be measured in several ways. 
The most straightforward method consists in measuring tour-to-tour
fluctuations. Since subsequent tours are statistically strictly uncorrelated, 
this gives immediately estimates for the error bars of any measured 
observable. In practice, not to waste CPU time, we bunched tours in 
groups of typically $10^4$ tours, and measured average values and 
fluctuations over these bunches.

\begin{figure}
\centerline{\psfig{file=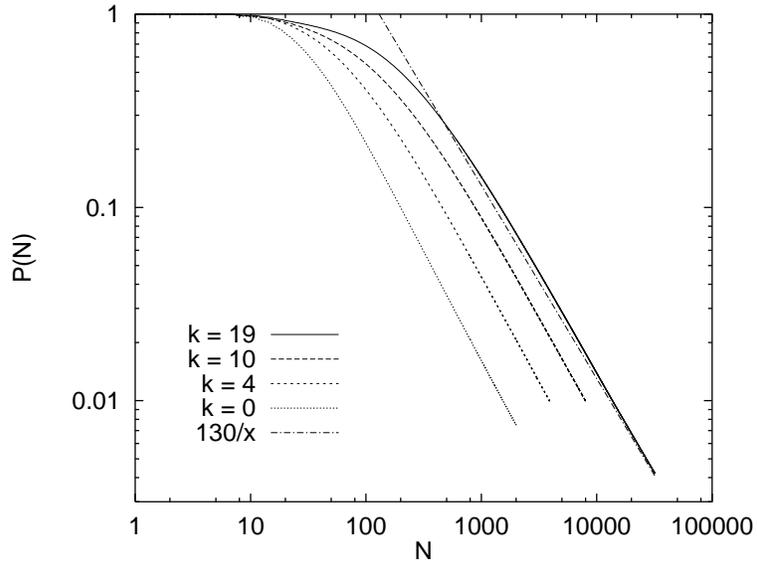,width=10.cm,angle=270}}
\caption{
Probabilities to reach length $N$ at least once in a tour, plotted against $N$. 
The curves are for PERM with $k$-step Markovian anticipation, with $k=19,10,5$, 
and $0$. The straight line is $130/N$.  }
\label{fig1}
\end{figure}

A less direct but more instructive measure of efficiency is the number 
of tours in which a length $N$ is reached in at least one configuration, 
normalized to the number of all tours. 
For other modified 
chain-growth algorithms such as incomplete enumeration \cite{redner} or 
the Berretti-Sokal (BS) algorithm \cite{beretti}, it seems that this number 
decreases as $const/N$, provided the cloning and pruning parameters are 
adjusted so that the total number of generated walks is independent of $N$.
Improving the details of the algorithm cannot change this scaling law, but 
can change the constant considerably. For incomplete enumeration and the 
BS algorithm, the constant is $O(1)$. For PERM without Markovian anticipation 
$const\approx 10$. For Markovian anticipation 
with $k=19$ we found $const \approx 130$, see fig. \ref{fig1}.

\begin{figure}
\centerline{\psfig{file=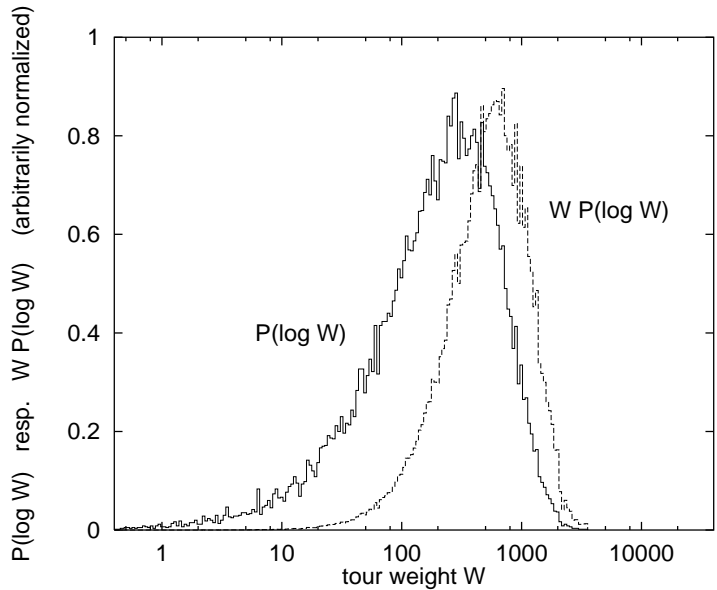,width=10.cm,angle=270}}
\caption{
Full line: histogram $P(\log W)$ of tours with fixed weight $W$, on a
logarithmic horizontal scale.
Normalization is arbitrary. Broken line: $W \times P(\log W) $,
again with arbitrary normalization.  }
\label{fig2}
\end{figure}

A last check for efficiency is the following. As we said, the weights of tours 
can vary substantially, to the point that the distribution of these weights 
might be not properly sampled in the region which should dominate 
statistical averages. If this happens, the estimates of the free energies 
are too low, and all the errors are systematically underestimated.
To show that this is not the case in the present simulation, we show in 
fig. \ref{fig2} the measured distribution of tour weights for $N=32000$, 
together with the weighted distribution. We see that the maximum 
of the weighted distribution occurs at a weight which is still well within
our sampling distribution.

%{\bf We can probably give here some data on the PERM simulation. 
%It could be organized in Table 1, together with the join-and-cut data.
%We could report: \# of starts, \# of successful tours, CPU-time 
%per successful tour (are these the relevant quantities, corresponding 
%to \# independent walks and tau in CPU-units for the j.and c.?) 
%It would be nice to have it on the same machine
%used to benchmark the join-and-cut. We could report total CPU-time and 
%see how the CPU-time per successful tour scales. The guess is $N^2$???}

\subsection{Determination of $\gamma$ using PERM data}

The raw data produced by PERM are estimates of the partition sum $z_N$, 
for all 
$N\le N_{\rm max}$. It is then straightforward to obtain 
estimates of $\gamma$.  Assuming Eq.~(\ref{cN}), 
the exponent $\gamma$ can be 
estimated from ratios of $z_{N_i}$ for three  different values of $N_i$. 
Indeed, if one chooses two 
positive constants $a < 1 <b$, and defines
\be
   \gamma_{\rm eff}(N) = {x\log z_{aN} + y\log z_{bN} - \log z_N \over 
           x \log a + y \log b } \;,                 \label{triple-1}
\ee
one checks easily that $\gamma_{\rm eff}(N) = \gamma$, 
provided that  \cite{sutter}
\be
   x+y =1 \;,\qquad ax+by=1 \;.                    \label{triple-2}
\ee 
If Eq.~(\ref{cN}) is not fulfilled exactly, Eqs.~(\ref{triple-1}) and 
(\ref{triple-2}) define effective $N$-dependent exponents. 
These exponents still 
depend on $a$ and $b$. It is clear that statistical errors will be small if 
$a\ll 1\ll b$, but then a wide range of chain lengths are needed. Since 
$N_{\rm max}$ is fixed and cannot be exceeded, this means that $aN$ will be 
small, and systematic errors will become large if we want to reduce statistical 
errors. In addition to the ratio $b/a$, one has the product $a\times b$ at one's disposal. 
For fixed $b/a$, its optimal value depends on the way how the statistical 
error on $z_N$ increases with $N$. 
In the present case, they increase roughly as $\sqrt{N}$. For practical 
reasons one also wants $a^{-1}$ and $b$ to be simple numbers. We found that 
good results were obtained with $a=1/2, b=4$ (and therefore $x=6/7, y=1/7$). 

The computation of the error bars is not trivial.
Indeed one should take into account the fact that different $z_N$ 
are correlated and therefore compute the full covariance matrix, 
which is practically impossible.
Therefore we proceeded differently to obtain error estimates.
For each bunch of tours we 
estimated $[\gamma_{\rm eff}(N)]_{\rm bunch}$ according to 
Eq.~(\ref{triple-1}), and from these 
values we obtained the variance of the latter. Notice that we did {\it not} use 
this procedure to obtain $\gamma_{\rm eff}(N)$ itself by taking 
the average over 
all bunches, as this would introduce a bias. 

In addition,
in order to improve the estimates of $X^{\rm cens}(N_{\rm min})$,
cf. Eq. \reff{Xcens}, obtained using the join-and-cut algorithm,
we evaluated this quantity using the PERM estimates of $z_N$. 
The estimate of the mean value is trivial. For the error bars,
we evaluated $[X^{\rm cens}(N_{\rm min})]_{\rm bunch}$ over bunches of tours
and then estimated their variance. We used the same trick as for
$\gamma_{\rm eff}$. Again, notice that
we did not use this procedure to estimate the mean value itself, 
since this introduces a bias. 
We have checked however that this bias
is extremely small.

\subsection{Dynamic results}

Let us now discuss the simulations using PERM.
We have performed high-statistics runs using Markovian 
anticipation with $k=19$. The dynamic data of the runs are reported
in Table \ref{table_perm_statistics}. The total PERM simulation
would have taken approximately
7 months of CPU time on an Alpha-Station 600 Mod 5/266.

We have first of all performed an analysis of the CPU time per 
independent configuration, which we expect to scale as 
$a N^z$ with $z\approx 2$. Fitting the data of Table 
\ref{table_perm_statistics}, and including in each fit only the data 
with $N\ge N_{\rm min}$ in order to detect systematic effects,
we have
\begin{eqnarray}
z &\approx& \cases{ 1.92    & \qquad \qquad  $N_{\rm min} = 500$, \cr
                    2.08    & \qquad\qquad   $N_{\rm min} = 2000$, }
\\
a &\approx& \cases{ 0.00003    & \qquad\qquad $N_{\rm min} = 500$, \cr
                    0.00001    & \qquad\qquad $N_{\rm min} = 2000$. }
\end{eqnarray}
The expected exponent $z\approx 2$ is clearly recovered. 
Let us now compare these results with the join-and-cut estimates. 
For the latter algorithm, the CPU time per independent configuration
scales as (cf. tables \ref{table_iter} and \ref{table_tau}) 
$\approx 0.0003 N^{1.6}$ and therefore the join-and-cut is faster 
in generating one independent configuration than the PERM 
as long as $N\gtapprox 1000$. This is may not be however a 
fair comparison of the 
two algorithms. Indeed what we call an ``independent configuration" 
for PERM is a bunch of walks which are correlated, but which still contain
more information than a single walk. In order to make a fair 
comparison it is better to consider an observable and compare the 
CPU time needed to obtain the same error bars. We have therefore
compared the results for our preferred observable
$X^{\rm cens}(N_{\rm min})$, see table \ref{tabella_valorimedi}.
We find that for $N_{\rm tot}=32000$ both algorithms require essentially
the same CPU time to produce data with the same error bars.
Only for longer walks the join-and-cut algorithm is more efficient.
Notice that this result is true if we use the Markovian anticipation.
In the absence of this improvement, PERM would be about ten times slower
and therefore less efficient than the join-and-cut algorithm already 
for $N\approx 500$.

\section{Estimate of $\gamma$}

Let us now discuss the results for the critical exponent $\gamma$.
We will first analyze the data using the method presented in 
section \ref{sec2.2}.
This will allow us to use at the same time the data produced using the 
join-and-cut and the PERM.
In table \ref{tabella_valorimedi} we report,
for various $N_{\rm min}$, 
the estimates of $X^{\rm cens}_{\rm MC} (N_{\rm min})$ obtained using the
two algorithms. We have checked that the raw data 
agree within error bars, therefore checking the 
correctness of our programs and error bars. 

\begin{figure}
\centerline{\psfig{file=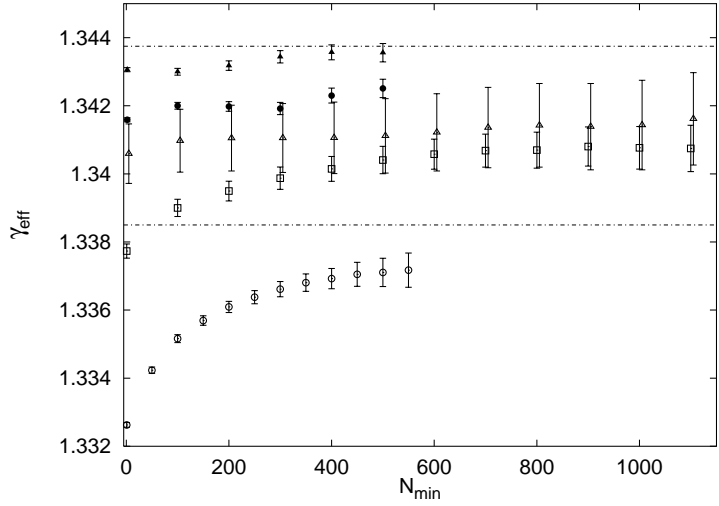,width=10.cm,angle=270}}
\caption{
Estimates of the effective exponents $\widehat{\gamma}$ for
$N_{\rm tot}=2000$ (empty circle), 
$8000$ (empty square), $32000$ (empty triangle) and
estimates obtained with the optimal amplitude method
for $\Delta=11/16$ (filled triangle) and $\Delta = 1$ (filled circle).
The dotted lines represent $\gamma_{\rm reg} = 43/32$  and the 
estimate of Ref. \protect\cite{Manhattan_Cardy}, $\gamma = 1.3385$.}
\label{fig3}
\end{figure}

%Questa e` la tabella
%     x              y             std. dev.(y)        weight
%     500.00000000000    4.9850700000000D-03    9.8797900000000D-12
%     2000.0000000000    7.0526900000000D-02    4.9107700000000D-08
%     4000.0000000000   0.29856600000000    9.7492100000000D-07
%     8000.0000000000    1.2726500000000    4.5427700000000D-04
%     32000.000000000    24.447400000000   0.71998500000000
%
%#pts
%  H =   0.
% 500.000&   1.91652&(  0.00000)&-17.21174&(  0.00000)&********&(       3 DF,&  0.00000%)\
%2000.000&   2.08181&(  0.00000)&-18.47537&(  0.00004)& 376.871&(       2 DF,&  0.00000%)\
%4000.000&   2.09175&(  0.00051)&-18.55784&(  0.00427)&   3.571&(       1 DF,&  5.87882%)\
%8000.000&   2.13189&(  0.02125)&-18.91858&(  0.19094)&   0.000&(       0 DF,&NaN      %)\
%
In table \ref{tabella_gamma} we report the estimates 
of the effective exponents $\widehat{\gamma}$ defined in Sec.~\ref{sec2.2}.
A graph of these estimates together with the results of the fits described 
below is reported in fig. \ref{fig3}.
It is evident that the corrections to scaling
are particularly strong and indeed $\widehat{\gamma}$ clearly increases 
with $N_{\rm min}$ and $N_{\rm tot}$. Under the only assumption that in the 
interval $1000 \ltapprox N \le 32000$ the corrections to scaling 
have the asymptotic sign, i.e. if we exclude that $\widehat{\gamma}$ 
will decrease for larger values of $N_{\rm min}$ and $N_{\rm tot}$, we obtain 
(using $N_{\rm tot}=32000$, $N_{\rm min}=2000$)
\be
{\gamma} \gtapprox 1.3423 \pm 0.0018\,.
\ee
This clearly excludes the result of Ref. \cite{Manhattan_Cardy}. Under
this  very weak assumption, we can conclude that there is no evidence from our 
data of the non-universality predicted in Ref. \cite{Cardy94}. 
More conservatively, our data indicate that, if the effect is really there,
it is extremely small: 
\be
\gamma_{\rm reg} - \gamma < 0.0014 \pm 0.0018\,.
\ee

We can improve this bound making additional assumptions on the corrections 
to scaling.  For each value of $N_{\rm min}$ we assume that
$z_N$ is well approximated by
\be
z_N = A \mu^N N^{\gamma-1} \left(1 + a N^{-\Delta}\right)
\ee
for $N\ge N_{\rm min}$.
We are not able to determine $\Delta$ reliably and therefore we have 
performed fits for various fixed values of $\Delta$.
Our analysis works as follows.
For each triple $(N_{\rm min},N_{\rm tot},a)$ and for each value of $\Delta$,
we can define an effective exponent
$\widehat{\gamma}(N_{\rm min},N_{\rm tot},a)$ (from now on we suppress the dependence
on $\Delta$) using the natural 
generalization of Eqs. \reff{eq_max_like} and \reff{mean-value-theor} with 
a corresponding error $\Delta\widehat{\gamma}(N_{\rm min},N_{\rm tot},a)$.
Then for each $N_{\rm min}$ and $\Delta$ we determine the optimal 
value $a_{\rm opt}$ of $a$ and an 
estimate of the exponent $\overline \gamma(N_{\rm min})$,
by minimizing with respect to $a$ and $\overline \gamma(N_{\rm min})$ 
\be
\chi^2 =\; \sum_{N_{\rm tot}}
\left[
      {\widehat{\gamma}(N_{\rm min},N_{\rm tot},a) - \overline{\gamma}(N_{\rm min})
      \over \Delta \widehat \gamma(N_{\rm min},N_{\rm tot},a)}
\right]^2 \,,
\ee 
where the sum runs over $N_{\rm tot}=2000$, $8000$, $32000$.
The statistical errors are obtained in a standard fashion.

In table \ref{tabella_estrapolazioni_gamma} we report the results of the
fits for various values of $N_{\rm min}$ and for two values of $\Delta$,
$\Delta = 1$ and $\Delta = 11/16$. 
First let us observe that the estimate of $\gamma$ decreases with 
increasing $\Delta$. Therefore a lower bound independent of any 
assumption on the value of $\Delta$ is obtained using 
the results for $\Delta = 1$. 
{}From  table \ref{tabella_estrapolazioni_gamma} we can estimate
\be
\gamma \gtapprox 1.3425 \pm 0.0003\,.
\label{lowerbound}
\ee
We can try to do more and see if we can obtain from the data 
a rough indication of the value of $\Delta$.  In principle we can consider 
the results for various $N_{\rm min}$ and see for which value of 
$\Delta$ the estimates do not depend on $N_{\rm min}$. Notice however
that the results with different $N_{\rm min}$ are strongly correlated 
--- they are obtained essentially from the same data --- and therefore
an observed stability or upward trend should be interpreted with caution 
since it could simply be an effect of the 
correlations. If we consider the results of table
\ref{tabella_estrapolazioni_gamma} we see that the analysis 
with $\Delta = 11/16$ is extremely stable: $a_{\rm opt}$ and $\gamma$ 
are essentially constant and we can estimate
\be
\gamma=1.3436 \pm 0.0003\,.
\ee
This result is in perfect agreement with the universal value 
$\gamma_{\rm reg} = 43/32$.  
On the other hand the results for $\Delta = 1$ show an upward 
systematic trend. However the effect is barely statistically 
significant (the results for $N_{\rm min} = 100$ and $N_{\rm min} = 500$ 
differ approximately by 1.5 combined error bars) and one could just
think that the systematic increase is simply an effect due 
to the neglected corrections to scaling and/or a result of the 
correlations. Without further hypotheses, we cannot confidently 
go beyond the lower bound \reff{lowerbound}. However if we assume 
$\gamma = \gamma_{\rm reg}$, then $\Delta = 1$ becomes less plausible and 
we can say that our data favour the presence of a non-analytic exponent.
We cannot give a reliable estimate of $\Delta$, but Saleur's value 
$11/16$ fits our data very well. 

\begin{figure}
\centerline{\psfig{file=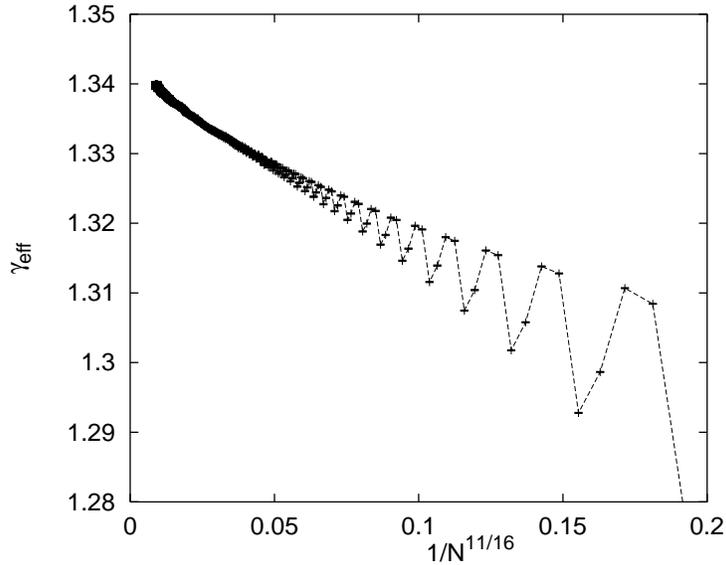,width=10.cm,angle=270}}
\caption{
Effective exponents $\gamma_{\rm eff}$ defined by Eq. (\ref{triple-1}), 
plotted against $N^{-11/16}$. The extrapolation to the $y$-axis 
gives the estimate of $\gamma$. Only data with high statistics, i.e.
$N < 1000 $ are included.}
\label{fig4}
\end{figure}

This conclusion is fully supported by the alternative analysis of the 
PERM data using Eq. (\ref{triple-1}). In fig.~\ref{fig4} we plot 
$\gamma_{\rm eff}$, obtained with $a=1/2, \,b=4$, against $N^{-11/16}$. 
If $\Delta$ has Saleur's value, and if there are no other corrections to 
scaling, we should expect a straight line intersecting the $y$-axis at 
$\gamma$. The most dramatic deviations from a straight line are strong 
period-four oscillations, observed also in \cite{Manhattan_Cardy}. 
Similar period-four oscillations are observed for SAWs on the square lattice 
\cite{conway93}.
They correspond to a singularity of the grand canonical partition sum 
at $x=-1/\mu$ \cite{Guttmann-Whittington_78}.
In fig.~\ref{fig4} one may also observe
a slight curvature which might suggest that 
$\Delta<11/16$. However, a more careful analysis, using also different pairs 
$(a,b)$ and additional correction-to-scaling terms, suggests that 
this effect is not significant. In contrast, the fact that $\Delta<1$ seems
significant. Accepting for $\Delta$ a value between 1/2
and 0.7, we find again perfect agreement with $\gamma_{\rm reg} = 43/32$, 
while the estimate of \cite{Manhattan_Cardy} seems ruled out.

\section*{Acknowledgments}
\bigskip
We thank John Cardy, Tony Guttmann, Flavio Seno, and 
Antonio Trovato for very helpful 
discussions.

\clearpage

\begin{table}
%\protect\footnotesize
\begin{center}
\begin{tabular}{|c|c|c|}
\hline
$ N_{\rm tot} $ & $N_{\rm iter}$
  & CPU-time \\
\hline\hline
500 & $10^9$ & 0.248(1)   \\
\hline
2000 & $12 \cdot 10^8$ & 0.846(8)    \\
\hline
8000 & $33 \cdot 10^8$ & 2.41(3)    \\
\hline
32000 & $23 \cdot 10^8$& 9.08(2)\\
\hline
\end{tabular}
\end{center}
\caption{Number of iterations and CPU times per iteration for various values
of $N_{\rm tot}$. CPU-times are expressed in ms and refer to an Alpha-Station 600 Mod 5/266.}
\label{table_iter}
\end{table}

\begin{table}
%\protect\footnotesize
\begin{center}
\begin{tabular}{|c|cc|cc|}
\hline
$ N_{\rm tot} $ & $f_{\rm piv}$  & $f_{\rm jc}$
  & $ \tau_{{\rm int},Y(1)}$
  & $ \tau_{{\rm int},Y(100)}$ \\
\hline\hline
500 & 0.451943(11) & 0.155468(11) & $4.912 \pm 0.031$& $8.171 \pm 0.023$\\
\hline
2000   & 0.3446316(86) & 0.0998629(77) &$14.30 \pm 0.23 $ & $19.38 \pm 0.12$\\
\hline
8000   & 0.2624050(83)& 0.0629925(65) & $49.6 \pm 1.2$ & $55.8 \pm 1.0$\\
\hline
32000  & 0.199693(46) & 0.039330(31)& $202.4 \pm 7.4$ & $206.7 \pm 7.0$\\
\hline
\end{tabular}
\end{center}
\caption{Acceptance fraction for the pivot move ($f_{\rm piv}$),
for the join-and-cut move ($f_{\rm jc}$)  and autocorrelation times
for the various values
of $N_{\rm tot}$. Autocorrelation times are expressed in units of two iterations.
}
\label{table_tau}
\end{table}

%-------------------------------------------------------

\begin{table}
\begin{center}
\begin{tabular}{|c|c|c|c|c|}
\hline
$ N_{\rm max} $ &    $N_{\rm tour}$     &    $N_{\rm config}$   &    $N_{\rm indep}$    &  CPU time \\ \hline\hline
            500 & $1.0\cdot 10^6$   & $1.140\cdot 10^6$ & $2.637\cdot 10^5$ &    4.985 \\ \hline
           2000 & $8.876\cdot 10^8$ & $1.010\cdot 10^9$ & $6.533\cdot 10^7$ &   70.53 \\ \hline
%                                                                                            SPARC 20, 296 MHz
           4000 & $1.756\cdot 10^8$ & $2.016\cdot 10^8$ & $6.359\cdot 10^6$ &  298.6 \\ \hline
           8000 & $3.421\cdot 10^8$ & $4.035\cdot 10^8$ & $6.296\cdot 10^6$ &  1272.6(4) \\ \hline     
%                                                                                            wpta15
          32000 & $2.792\cdot 10^7$ & $3.703\cdot 10^7$ & $1.154\cdot 10^5$ & $ 244(7)\cdot 10^2$ \\ \hline
\end{tabular}
\end{center}
\caption{Number of PERM tours (column 2) 
with $N\leq N_{\rm max}$, total number of
configurations with $N=N_{\rm max}$ (column 3), 
number of independent configurations with
$N=N_{\rm max}$ (i.e., of tours which reached 
$N=N_{\rm max}$; column 4), and CPU time per
independent configuration (expressed in ms, on an Alpha-Station 600 Mod 5/266;
column 5).}
\label{table_perm_statistics}
\end{table}

%-------------------------------------------------------

\begin{table}
\protect\footnotesize
\begin{center}
\begin{tabular}{|c|c|c||c|c|c|}
\hline
   & \multicolumn{2}{|c||}{$N_{\rm tot} = 500$} &
   & \multicolumn{2}{|c|}{$N_{\rm tot} = 2000$} \\
\hline
$ N_{\rm min} $ & $ X^{\rm cens}_{\rm PERM} $ & $ X^{\rm cens}_{\rm jc}$ & 
$ N_{\rm min} $ & $ X^{\rm cens}_{\rm PERM} $ & $ X^{\rm cens}_{\rm jc}$ \\
\hline
    2 & $10.59979820 (1214)$ & $10.5997734   (851) $&     2 & $13.3738441 (259)$ & $ 13.374033   (132)$  \\
   20 & $10.68281651 (924)$ & $10.6828003   (692) $&    50 & $13.4285517 (215)$ & $ 13.428669   (115)$  \\
   40 & $10.75716591 (695)$ & $10.7571316   (555) $&   100 & $13.4796982 (176)$ & $ 13.479777   (100)$  \\
   60 & $10.81801510 (514)$ & $10.8179770   (447) $&   150 & $13.5246592 (144)$ & $ 13.524727   (87)$  \\
   80 & $10.86850452 (365)$ & $10.8684807   (356) $&   200 & $13.5644369 (118)$ & $ 13.564528   (77)$  \\
  100 & $10.91057682 (248)$ & $10.9105787   (280) $&   250 & $13.5998096 (96)$ & $ 13.599867   (67)$  \\
  120 & $10.94553280 (170)$ & $10.9455605   (216) $&   300 & $13.6313610 (77)$ & $ 13.631408   (58)$  \\
  140 & $10.97427940 (116)$ & $10.9742889   (162) $&   350 & $13.6595411 (61)$ & $ 13.659569   (51)$  \\
  160 & $10.99746609 (70)$ & $10.9974737   (115) $&   400 & $13.6847010 (48)$ & $ 13.684742   (43)$  \\
  180 & $11.01557064 (39)$ & $11.0155716   (77) $&   450 & $13.7071218 (38)$ & $ 13.707140   (37)$  \\
  200 & $11.02893162 (21)$ & $11.0289259   (46) $&   500 & $13.7270279 (29)$ & $ 13.727030   (31)$  \\
\hline
\hline
   & \multicolumn{2}{|c||}{$N_{\rm tot} = 8000$} &
   & \multicolumn{2}{|c|}{$N_{\rm tot} = 32000$} \\
\hline
$ N_{\rm min} $ & $ X^{\rm cens}_{\rm PERM} $ & $ X^{\rm cens}_{\rm jc}$ & 
$ N_{\rm min} $ & $ X^{\rm cens}_{\rm PERM} $ & $ X^{\rm cens}_{\rm jc}$ \\
\hline
    2 & $ 16.148415 (85) $ & $ 16.148470   (199 ) $&     2 & $  18.92202 (76) $ & $  18.92211   (36) $  \\
  100 & $ 16.175459 (77) $ & $ 16.175516   (184 ) $&   100 & $  18.92796 (75) $ & $  18.92804   (35) $  \\
  200 & $ 16.203131 (70) $ & $ 16.203156   (171 ) $&   200 & $  18.93483 (73) $ & $  18.93486   (34) $  \\
  300 & $ 16.229231 (64) $ & $ 16.229317   (160 ) $&   300 & $  18.94185 (71) $ & $  18.94184   (33) $  \\
  400 & $ 16.253729 (58) $ & $ 16.253832   (150 ) $&   400 & $  18.94889 (70) $ & $  18.94884   (33) $  \\
  500 & $ 16.276733 (52) $ & $ 16.276882   (140 ) $&   500 & $  18.95588 (68) $ & $  18.95582   (32) $  \\
  600 & $ 16.298361 (48) $ & $ 16.298492   (131 ) $&   600 & $  18.96280 (66) $ & $  18.96275   (31) $  \\
  700 & $ 16.318720 (43) $ & $ 16.318815   (123 ) $&   700 & $  18.96964 (65) $ & $  18.96960   (31) $  \\
  800 & $ 16.337910 (39) $ & $ 16.337938   (114 ) $&   800 & $  18.97639 (63) $ & $  18.97635   (30) $  \\
  900 & $ 16.356014 (36) $ & $ 16.356073   (106 ) $&   900 & $  18.98303 (62) $ & $  18.98297   (30) $  \\
 1000 & $ 16.373112 (32) $ & $ 16.373133   (100 ) $&  1000 & $  18.98958 (60) $ & $  18.98951   (29) $  \\
 1100 & $ 16.389269 (29) $ & $ 16.389252   (93 ) $&  1100 & $  18.99603 (59) $ & $  18.99599   (29) $  \\
 1200 & $ 16.404539 (26) $ & $ 16.404505   (87 ) $&  1200 & $  19.00237 (58) $ & $  19.00235   (28) $  \\
 1300 & $ 16.418976 (24) $ & $ 16.418930   (81 ) $&  1300 & $  19.00863 (56) $ & $  19.00853   (27) $  \\
 1500 & $ 16.445524 (19) $ & $ 16.445475   (71 ) $&  1500 & $  19.02085 (53) $ & $  19.02071   (26) $  \\
& & &  2000& $  19.04978 (47) $ & $  19.04973   (24)$  \\
& & &  2500& $  19.07657 (42) $ & $  19.07649   (22)$  \\
& & &  3000& $  19.10144 (37) $ & $  19.10130   (21)$  \\
\hline
\end{tabular}
\end{center}
\caption{Raw data for $X^{\rm cens}_{\rm MC}$. We report separately the results obtained with the PERM and join-and-cut algorithms.} 
\label{tabella_valorimedi}
\end{table}

\begin{table}
\protect\footnotesize
\begin{center}
\begin{tabular}{|c|c||c|c||c|c||c|c|}
\hline
$N_{\rm min}$ & $ \gamma(500)$ & $N_{\rm min}$ & $ \gamma(2000) $ & $N_{\rm min}$ & $ \gamma(8000) $ & $N_{\rm min}$ & $ \gamma(32000) $ \\
\hline
    2 &   1.32950 (3)  &     2 &   1.33263 (7)  &     2 &    1.33773 (21)  &     2 &    1.34059 (87)  \\
   20 &   1.32584 (5)  &    50 &   1.33423 (9)  &   100 &    1.33900 (25)  &   100 &    1.34097 (92)  \\
   40 &   1.32776 (7)  &   100 &   1.33516 (12)  &   200 &    1.33950 (29)  &   200 &    1.34105 (97)  \\
   60 &   1.32880 (9)  &   150 &   1.33569 (14)  &   300 &    1.33987 (33)  &   300 &    1.34105 (101)  \\
   80 &   1.32955 (12)  &   200 &   1.33609 (17)  &   400 &    1.34015 (36)  &   400 &    1.34106 (105)  \\
  100 &   1.33011 (15)  &   250 &   1.33638 (19)  &   500 &    1.34041 (40)  &   500 &    1.34112 (109)  \\
  120 &   1.33064 (20)  &   300 &   1.33661 (22)  &   600 &    1.34058 (44)  &   600 &    1.34122 (114)  \\
  140 &   1.33120 (28)  &   350 &   1.33681 (26)  &   700 &    1.34068 (48)  &   700 &    1.34136 (118)  \\
  160 &   1.33139 (40)  &   400 &   1.33692 (30)  &   800 &    1.34070 (53)  &   800 &    1.34143 (123)  \\
  180 &   1.33051 (63)  &   450 &   1.33705 (35)  &   900 &    1.34080 (58)  &   900 &    1.34139 (127)  \\
  200 &   1.33235 (132)  &   500 &   1.33711 (42)  &  1000 &    1.34077 (63)  &  1000 &    1.34143 (131)  \\
    &                          &     &                          &  1100 &   1.34075 (68)  &  1100 &    1.34162 (136)  \\
    &                          &     &                          &  1200 &   1.34073 (74)  &  1200 &    1.34172 (139)  \\
    &                          &     &                          &  1300 &   1.34072 (80)  &  1300 &    1.34147 (143)  \\
    &                          &     &                          &  1500 &   1.34077 (94)  &  1500 &    1.34140 (153)  \\
    &                          &     &                          &     &                          &  2000 &    1.34232 (178)  \\
    &                          &     &                          &     &                          &  2500 &    1.34256 (205)  \\
    &                          &     &                          &     &                          &  3000 &    1.34250 (236)  \\
\hline
\end{tabular}
\end{center}
\caption{Estimates of $\gamma(N_{\rm tot})$ for various values of $N_{\rm min}$ and
$N_{\rm tot}$. }
\label{tabella_gamma}
\end{table}

\begin{table}
\protect\footnotesize
\begin{center}
\begin{tabular}{|c|cc|cc|}
\hline
$N_{\rm min}$  & \multicolumn{2}{|c|}{$\Delta=11/16$} 
           & \multicolumn{2}{c|}{$\Delta=1$} \\
\hline
   & $\gamma$ & $a_{\rm opt}$ & $\gamma$ & $a_{\rm opt}$ \\
\hline
    2  &     1.34306 (6)  &   0.43 (1) &
             1.34159 (6)  &   0.85 (2) \\
    100  &     1.34300 (10)  &   0.51 (4) &
               1.34200 (10)  &   1.75 (13) \\
    200  &     1.34318 (14)  &   0.54 (4) &
               1.34198 (14)  &   1.93 (24) \\
    300  &     1.34344 (18)  &   0.57 (5) &
               1.34192 (18)  &   1.98 (36) \\
    400  &     1.34357 (22)  &   0.59 (5) &
               1.34230 (22)  &   2.23 (50) \\
    500  &     1.34356 (27)  &   0.59 (6) &
               1.34251 (27)  &   2.40 (59) \\
\hline
\end{tabular}
\end{center}
\caption{Extrapolated estimates of $\gamma$ 
for $\Delta=11/16$ and $\Delta=1$.}
\label{tabella_estrapolazioni_gamma}
\end{table}

\eject

% \section*{Figure Captions}

% \begin{figure}
% \begin{center}
% \epsfxsize = 0.9\textwidth
% \leavevmode\epsffile{tours.ps}
% \end{center}
% \caption{
% Probabilities to reach length $N$ at least once in a tour, plotted against $N$. 
% The curves are for PERM with $k$-step markovian anticipation, with $k=19,10,5$, 
% and $0$. The straight line is $130/N$.
% }
% \label{fig1}
% \end{figure}

% \begin{figure}
% \vspace*{-1cm} \hspace*{-0cm}
% \begin{center}
% \epsfxsize = 0.9\textwidth
% %%\leavevmode\epsffile{tour-weights.ps}
% \quad \vspace{1in}  %% TEMPORARY UNTIL FILE IS THERE
% \end{center}
% \vspace*{-1cm}
% \caption{
% Full line: histogram $P(log W)$ of tours with fixed weight $W$, on a 
% logarithmic scale. 
% Normalization is arbitrary. Broken line: $W \times P(log W) $, 
% again with arbitrary normalization.
% }
% \label{fig2}
% \end{figure}

\end{document}